\documentclass[twocolumn,preprintnumbers,amsmath,amssymb]{revtex4}

\usepackage{graphicx}
\usepackage{dcolumn}
\usepackage{bm}
\usepackage{epsfig}
\usepackage{amssymb}
\begin{document}

\preprint{}

\title{Comments on arXiv:1006.0972 ``XENON10/100 dark matter constraints in 
comparison with CoGeNT and DAMA: examining the 
$\mathcal{L}_{\text{eff}}$ dependence''}

\author{J.I. 
Collar$^{a}$
}
\address{ 
$^{a}$Enrico 
Fermi Institute, Kavli Institute for Cosmological Physics and  Department of Physics, University of Chicago, Chicago, IL 60637\\
}
\maketitle

Savage {\it et al.} \cite{chris} have recently put forward the claim that results from 
the XENON10 experiment are incompatible with the totality of both 
DAMA/LIBRA \cite{DAMA} and 
CoGeNT \cite{cogent} experimental regions. In this brief note the source of this 
erroneous conclusion is identified in a misinterpretation of the 
XENON10 efficiency in the detection of {\it S}1 light from low-energy nuclear recoils. 

I intend to keep this discussion factual and brief and refer an interested 
reader to the literature cited: the main thesis in \cite{chris} is that 
XENON10 provides more astringent light-mass WIMP limits than 
XENON100, by virtue of its lower energy threshold (2 {\it S}1 photoelectrons 
(PE) or $\sim$4.6 keV$_{r}$ as opposed to 4 {\it S}1 PE or $\sim$9.5 keV$_{r}$). Unfortunately, 
from comments made in \cite{chris}, it is clear that the authors have 
neglected to include in their analysis the effect of the XENON10 efficiency in 
extracting a larger than two-fold {\it S}1 PE coincidence, one of the 
requirements for acceptance of an event in both XENON10 and XENON100. 

This efficiency is 
assumed by Savage {\it et al.} to be a flat 100\% down to zero recoil 
energy in XENON10. 
In reality it is a rapidly decreasing 
function below $\sim$10 keV$_{r}$, becoming zero somewhere in the 
region 1 keV$_{r}$ to 3.5 keV$_{r}$ $^{1}$\footnotetext[1]{Here a cutoff value of 2 keV$_{r}$ 
corresponding to the Manzur {\it et al.} $\mathcal{L}_{\text{eff}}$ 
model is adopted for consistency.}. The 
importance of this $\it S$1 detection 
efficiency is 
mentioned in a recent release from XENON100 \cite{xenon100} (by 
remarking that it is a relatively minor issue when dealing with a 4 {\it S}1
PE or $\sim$9.5 
keV$_{r}$ threshold in that detector), is covered in technical detail for 
XENON10 in \cite{nim}, and has 
been recently emphasized in \cite{peter}. By adopting the 
same erroneous efficiency it is possible to obtain results (dotted green lines in 
top Fig.\ 1) essentially 
indistinguishable from those derived by Savage {\it et al.} 
Once the correct efficiency is included,  limits are extracted that are 
much more relaxed 
(rest of the lines in top Fig.\ 1), showing compatibility with 
both DAMA/LIBRA and CoGeNT, in particular when the additional sources 
of uncertainty discussed below are included. For clarity, only the 
lower 1 $\sigma$ C.L. boundary (dashed green line) for 
the most conservative $\mathcal{L}_{\text{eff}}$ extrapolation 
attempted by Savage {\it et al.} is included in Fig.\ 1. The solid 
green lines correspond to the dotted green lines when the correct $\it S$1 
efficiency is applied.

\begin{figure}
\includegraphics[width=8.5cm]{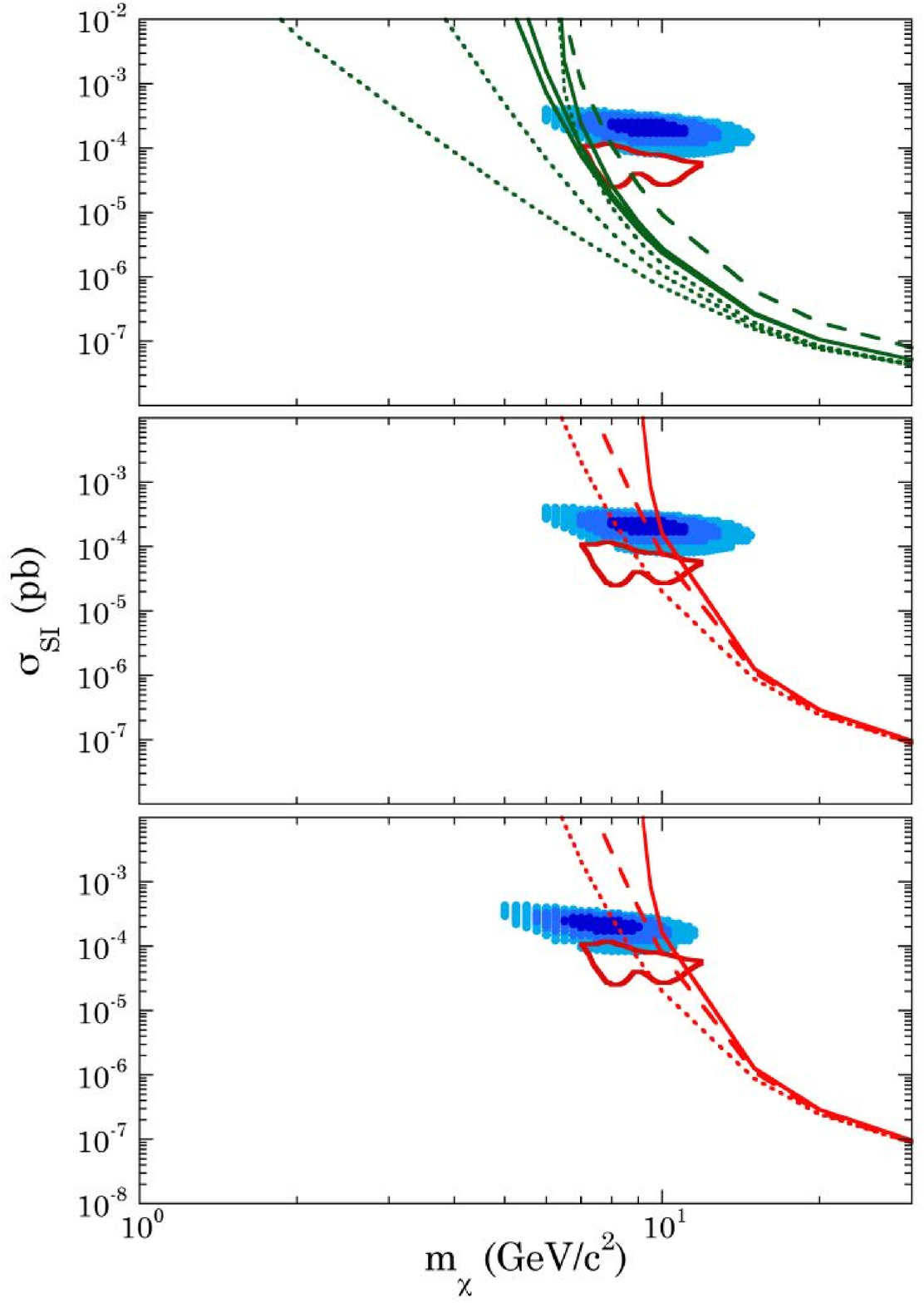}
\caption{\label{fig:epsart} XENON10 exclusion limits obtained under a variety 
of assumptions, some erroneous, some not (see text).
For consistency we use here our own derived DAMA/LIBRA and CoGeNT 
regions (1$\sigma$,2$\sigma$,3$\sigma$ and approximate 90\% C.L. 
respectively), employing the same astrophysical parameters as in 
\protect\cite{chris,smith}.
}
\end{figure}

While mistakes can be made and will be made, a
limited treatment of the uncertainties affecting these experiments 
is also evident in \cite{chris}. This is in contrast with the more 
balanced position that some of us 
have attempted to take in \cite{ours}. It is claimed in 
\cite{chris} that the most conservative XENON constraints can be 
obtained by attempting different low-energy extrapolations of the 
scintillation efficiency function $\mathcal{L}_{\text{eff}}$ derived 
by Manzur {\it et al.} \cite{dan1}. While $\mathcal{L}_{\text{eff}}$ for liquid 
xenon (LXe) should  
in principle be a detector-independent quantity (unfortunately far from 
experimental reality as of today), the detector-specific $\it S$1 detection 
efficiency mentioned above plays a role in its derivation 
\cite{nim,peter}. It would 
then seem natural to examine the sensitivity that can be obtained 
from XENON10 data by using the latest $\mathcal{L}_{\text{eff}}$ that 
can be derived from that very same experiment \cite{peter}. This is 
represented in Fig.\ 1 (middle panel) by a dotted red line. 
Sub-threshold Poisson fluctuations are conservatively assumed in all these 
calculations, as well as an allowance for a non-zero $\mathcal{L}_{\text{eff}}$ all 
the way down to zero recoil energy. For the case of XENON10, this is done via an adiabatic fit as in 
\cite{ahlen}, which yields good agreement with the expected kinematic 
cutoff for LXe \cite{ours}. This assumption of finite response down 
to zero recoil energy is another conservative premise that we have emphasized does not have to  
be warranted in reality, and one that dominates the sensitivity 
extracted for these low-mass WIMPs
\cite{ours}. Other values of $\mathcal{L}_{\text{eff}}$ neglected by 
Savage {\it et al.}, such as that recently measured by the ZEPLIN 
collaboration \cite{zeplin}, provide even more conservative limits 
(red dashed lines) under similar generous premises. It is observed 
that both $\mathcal{L}_{\text{eff}}$ functions derived from ZEPLIN 
and most recently for XENON10 are 
compatible within uncertainties with a zero value for 
$\mathcal{L}_{\text{eff}}$ below $\sim$8 
keV$_{r}$ once the $\it S$1 threshold bias is included for the second 
\cite{peter}. We also examine this most moderate possibility, obtaining 
limits (solid red lines) not very different from those that can be 
generated in the absence of Poisson fluctuations \cite{peter}. 

The bottom panel in Fig.\ 1 is offered as an illustration of the many 
additional uncertainties affecting dark matter experiments in this low-mass WIMP 
region, a subject already brought up in \cite{ours}. The position of the DAMA/LIBRA favored region is arguably 
particularly sensitive to these, since its derivation includes additional halo 
model uncertainties via the magnitude of the annual modulation. Here 
a simple minor alteration to the quenching factors for 
NaI[Tl] is made, in going from the usual Q$_{Na}$=0.3, Q$_{I}$=0.09 
applied in the top and middle panels, to a Q$_{Na}$=0.4, 
Q$_{I}$=0.05 as derived in \cite{quenching}. Just this small 
modification is enough to considerably relax any tension with XENON10 
constraints. Such additional uncertainties will be treated in an 
upcoming publication.

In conclusion, it is found that the main thesis in Savage {\it et al.}, 
namely that XENON10 ``is incompatible with the DAMA/LIBRA 3$\sigma$ region 
and the 7-12 GeV WIMP mass region of interest in CoGeNT'' is
the result of a mistake in their understanding the XENON10 
efficiencies, together with an additional incomplete treatment of present-day uncertainties. 
No attempt has been made here to cross-check any of the 
findings made by Savage {\it et al.} in respect to the other 
experiments they mention.  
An interested reader is referred to \cite{ours} for a careful treatment of 
XENON100. The intricacies of the
experimental details and abundant uncertainties affecting the 
low-mass WIMP region, specially in the case of 
LXe, do not invite any
rushed conclusions.

The author is indebted to G. Gelmini and D.N. McKinsey for several useful
comments.

\end{document}